\documentclass[conference]{IEEEtran}
\IEEEoverridecommandlockouts
\usepackage{amsmath,amssymb,amsfonts}
\usepackage{algorithmic}
\usepackage[ruled,vlined]{algorithm2e}
\usepackage{graphicx}
\usepackage{textcomp}
\usepackage{xcolor}
\usepackage{lipsum}  
\usepackage{subcaption}
\usepackage{booktabs}
\usepackage{verbatim}
\usepackage{censor}
\usepackage{lineno}

\usepackage[utf8]{inputenc}
\usepackage{pgfplots}
\pgfplotsset{compat=newest}
\usepgfplotslibrary{groupplots}
\usepgfplotslibrary{dateplot}
\usepackage{pstricks,auto-pst-pdf}
\usepackage[keeplastbox]{flushend}
\usepackage{tikz, pgfplots}
\pgfplotsset{compat=1.16}

\usepackage[style=ieee, url=false, doi=false, natbib=true, mincitenames=1, maxcitenames=1]{biblatex}

\usepackage[capitalize]{cleveref}

\definecolor{vandeusen}{RGB}{73, 92, 111}

\def\BibTeX{{\rm B\kern-.05em{\sc i\kern-.025em b}\kern-.08em
    T\kern-.1667em\lower.7ex\hbox{E}\kern-.125emX}}
\begin{document}

\title{Dyadic Sex Composition and Task Classification Using fNIRS Hyperscanning Data}

\author{
\IEEEauthorblockN{Liam A. Kruse}
\IEEEauthorblockA{\textit{Aeronautics and Astronautics} \\
\textit{Stanford University}\\
Stanford, CA 94305 \\
lkruse@stanford.edu} \\
\IEEEauthorblockN{Mykel J. Kochenderfer}
\IEEEauthorblockA{\textit{Aeronautics and Astronautics} \\
\textit{Stanford University}\\
Stanford, CA 94305 \\
mykel@stanford.edu}
\and

\IEEEauthorblockN{Allan L. Reiss}
\IEEEauthorblockA{\textit{Psychiatry and Behavioral Sciences} \\
\textit{Stanford University, School of Medicine}\\
Stanford, CA 94305 \\
areiss1@stanford.edu}\\
\IEEEauthorblockN{Stephanie Balters}
\IEEEauthorblockA{\textit{Psychiatry and Behavioral Sciences} \\
\textit{Stanford University, School of Medicine}\\
Stanford, CA 94305 \\
balters@stanford.edu}
}

\maketitle

\begin{abstract}
Hyperscanning with functional near-infrared spectroscopy (fNIRS) is an emerging neuroimaging application that measures the nuanced neural signatures underlying social interactions. Researchers have assessed the effect of sex and task type (e.g., cooperation versus competition) on inter-brain coherence during human-to-human interactions. However, no work has yet used deep learning-based approaches to extract insights into sex and task-based differences in an fNIRS hyperscanning context. This work proposes a convolutional neural network-based approach to dyadic sex composition and task classification for an extensive hyperscanning dataset with \textit{N} = 222 participants. Inter-brain signal similarity computed using dynamic time warping is used as the input data. The proposed approach achieves a maximum classification accuracy of greater than 80 percent, thereby providing a new avenue for exploring and understanding complex brain behavior.\\
\end{abstract}

\section{Introduction}
Understanding the neural and behavioral signatures of human social behavior in the presence of robotic systems is a topic of burgeoning interest in the applied robotics community. Significant theoretical and technological advances have been made in capturing and mediating interactions between machines and a single human actor. Examples of automated human-machine-loops include real-time driver drowsiness detection systems \cite{mehta2019real}, brain computer interface (BCI) systems that control assistive medical devices for persons with motor disabilities \cite{peng2018single}, and adaptive automation solutions to monitor air traffic controller attention and workload \cite{di2019brain}. However, the difficulty of measuring and mediating team dynamics increases considerably when additional human actors join the human-machine interaction to form complex human-human-machine (HHM) scenarios. Initial research thrusts have been conducted to explore HHM interactions such as robots working as emotion regulators to positively mediate conflict and instill trust in teams \cite{jung2015using, de2020towards}. The emergence of hyperscanning with portable functional near-infrared spectroscopy (fNIRS) neuroimaging could facilitate the systematic integration of social information during HHM interactions to the robotic systems of tomorrow. As a first step towards timely and accurate interpretation of neural signatures during social interactions, we present a novel approach to classify sex (i.e., female versus male) and task type (cooperation versus competition) and test the approach on an extensive fNIRS hyperscanning dataset with \textit{N} = 222 participants. 

Functional NIRS is a non-invasive neuroimaging technology that assesses cortical activity with relatively high spatial resolution compared to electroencephalography (EEG) \cite{scholkmann2014review} and relatively high temporal resolution compared to functional magnetic resonance imaging (fMRI) \cite{cui2011quantitative}. The technology uses near-infrared light to measure the hemodynamic response of the cerebral cortex as a proxy for neural activity \cite{peng2018single, ayaz2009assessment}. In recent years, fNIRS systems have become increasingly portable and affordable, allowing researchers to investigate neurocognitive behavior in naturalistic settings \cite{baker2017portable}.
Researchers have extended fNIRS measurements from single brain to hyperscanning applications to investigate shared brain functions related to social interactions in a laboratory \cite{cui2012nirs, funane2011synchronous} and increasingly naturalistic settings \cite{liu2016nirs,miller2019inter,mayseless2019real}. We refer the reader to \citet{balters2020capturing} for a more comprehensive introduction to fNIRS hyperscanning. Results from fNIRS hyperscanning studies have shown increased inter-brain coherence to be related to enhanced levels of interaction between team members \cite{cui2012nirs, liu2016nirs, miller2019inter, mayseless2019real, liu2017inter}. Furthermore, nuanced inter-brain coherence patterns emerge depending on the sex composition and task objectives of the interacting dyad. \Citet{baker2016sex} and \citet{cheng2015synchronous} conducted wavelet coherence analyses on fNIRS hyperscanning data for dyadic cooperation tasks and found that inter-brain coherence is highly dependent on the sex composition of the dyad. Sex differences in fNIRS neural signatures also emerge during spontaneous face-to-face deception \cite{zhang2017gender}, risky decision making during gambling games \cite{zhang2017social}, and group creative idea generation \cite{lu2020gender}. Other researchers identified significant differences in inter-brain coherence values between cooperation and competition tasks within the superior frontal cortex \cite{cui2012nirs} and right posterior superior temporal sulcus \cite{liu2017inter}.

Employing fNIRS hyperscanning technology to infer the identity and objective of human participants engaged in HHM interactions would provide tremendous utility in applied robotics scenarios.
For example, future pilot-copilot-autopilot systems could execute judicious control authority switches based on the timely interpretation of pilots' joint task attention. In this work, we conduct an initial foray into deciphering  \textit{who} is interacting in a given HHM scenario and \textit{how} they are engaging. We propose a novel approach to dyadic sex composition and task classification using a modified form of the ubiquitous LeNet-5 convolutional neural network (CNN) architecture proposed by \citet{lecun1998gradient}. We use inter-brain signal similarity computed using dynamic time warping (DTW) as the input data and validate our approach on an existing fNIRS hyperscanning data set \cite{baker2016sex} in which female-female and male-male dyads execute both cooperative and competitive tasks of within-subject design. To the best of the authors' knowledge, this is the first time that CNNs have been leveraged to classify hyperscanning data and one of the first applications of CNNs within the broader fNIRS community.

\section{Related Work} \label{Related}
Researchers have employed a multitude of machine learning techniques to classify fNIRS single-brain signals. \Citet{shamsi2019multi} implemented a support vector machine (SVM) with a quadratic polynomial kernel to classify movement execution tasks. \Citet{peng2018single} tested both a linear discriminant analysis algorithm and an SVM algorithm to classify motor imagery tasks wherein participants moved an on-screen object in their imagination. \Citet{power2010classification} employed a hidden Markov model to differentiate between mental arithmetic and music imagery, demonstrating the potential for a BCI device based on cognitive tasks rather than motor tasks. Hidden Markov models have also successfully identified finger-tapping tasks \cite{sitaram2007temporal} and detected music imagery \cite{falk2010improving} from fNIRS data. Still other researchers have conducted fNIRS classification tasks using k-nearest neighbors and Na\"{i}ve Bayes \cite{gottemukkula2011classification, ayaz2009assessment, shin2014multiclass}. \Citet{saadati2019convolutional} and  \citet{asgher2020classification} investigated the use of convolutional neural networks for mental workload classification; these remain the only instances of CNN-based classifiers for fNIRS data. 

Accurately classifying motor tasks and cognitive imagery is empirically challenging even in the simplest instances of binary classification. \Citet{shamsi2019multi} achieved an average accuracy of 70.43\%  and \citet{power2010classification} obtained an average accuracy of 77.20\% for binary classification problems, while \citet{shamsi2019multi} and \citet{peng2018single} achieved  accuracies of 78.55\% and 39.98\% for five and four-class problems, respectively. The two CNN-based approaches yield some of the highest accuracies for fNIRS classification tasks, with average four-class classification accuracies of 89.00\% by \citet{saadati2019convolutional} and 83.45\% by \citet{asgher2020classification}. The ability of CNNs to extract the most salient features with minimal levels of \textit{a priori} feature extraction likely contributes to the state-of-the-art classification accuracies and inspires the approach developed in this work. We propose a CNN architecture based on the LeNet-5 architecture to predict dyadic sex composition given knowledge of the task type and to predict the task type given knowledge of the dyadic sex composition.

\section{Hyperscanning Dataset and Features}

\subsection{Data Acquisition}
We used an extensive hyperscanning dataset that is partially introduced and analyzed in \citet{baker2016sex}. A total of 222 healthy adults (110 females, 112 males) were recruited for the hyperscanning study. All participants were right-handed with normal or corrected to normal vision and hearing. Each participant was paired with a random individual to form a dyad. The dataset contains 38 female-female, 34 female-male, and 39 male-male pairs.  Participants did not interact prior to the study and were not matched based on age or ethnicity \cite{baker2016sex}.

Each dyad performed a series of consecutive cooperation and competition tasks on computer screens while sitting on opposite sides of a table. The tasks were grouped into blocks of twenty trials and the task order was randomized across dyads to reduce bias. For cooperation trials, participants attempted to synchronize a button press event; for competition trials, participants raced to press a button before their partner. Button press responses were elicited from participants after an onscreen visual cue. The intertrial intervals were varied to reduce habituation bias.

Cortical hemodynamic activity of each participant was recorded using a continuous wave fNIRS system (ETG-4000, Hitachi, Japan) with a sampling frequency of 10 Hz. Fifteen optodes (8 sources × 7 detectors) were placed over the right inferior prefrontal cortex (PFC), right frontopolar PFC, right superior PFC, and right dorsal lateral PFC according to the international 10--20 EEG placement system, resulting in a total of 19 fNIRS channels. 

\subsection{Functional NIRS Preprocessing}
We used \textsc{Matlab}-based functions derived from \textsc{Homer 2} \cite{huppert2009homer} and followed preprocessing steps as described in \citet{baker2016sex}. We motion corrected optical density data using a wavelet-based motion correction filter \cite{molavi2012wavelet} and applied a bandpass filter (0.01--0.5 Hz) to eliminate systemic noise (e.g., cardiac and respiratory signals). The filtered optical density data was subsequently converted to oxy-hemoglobin (HbO) and deoxy-hemoglobin (HbR) values using the modified Beers-Lambert Law \cite{kocsis2006modified}. Since HbO measures are known to be more robust and sensitive to task-associated changes compared to HbR measures \cite{ferrari2012brief,plichta2006event}, we only used HbO data for further analysis. 

We eliminated one channel of HbO data due to poor signal quality, resulting in a total of 18 channels. Furthermore, we removed all female-male dyads, keeping only male-male and female-female dyadic sex compositions to simplify the classification task; binary classification is a typical first step in fNIRS classification studies \cite{peng2018single, power2010classification}. The fNIRS recorded optical density data at a frequency of 10 Hz; thus, a trial with 60 data points had a duration of 6.0 seconds. All trials containing fewer than 50 or more than 60 data points were removed from the dataset to eliminate trials with outlier intertrial intervals. To convert the data into a consistent input size for the proposed CNN-based approach, we subsequently trimmed the remaining time series to only consider the last 50 data points for each trial; establishing consistent timing intervals is common practice in fNIRS classification problems \cite{shamsi2019multi, peng2018single}.  Lastly, we scaled each sample to unit norm. A total of 3,188 trials remained at the conclusion of the data cleaning process; \cref{tab:dataset} summarizes the dataset. We consider both dimensions of the dataset because we seek to predict the dyadic sex composition given knowledge of the task type and to predict the task type given knowledge of the dyadic sex composition.

\begin{figure}[!htb]
\centering
\resizebox{0.4\textwidth}{!}{\input{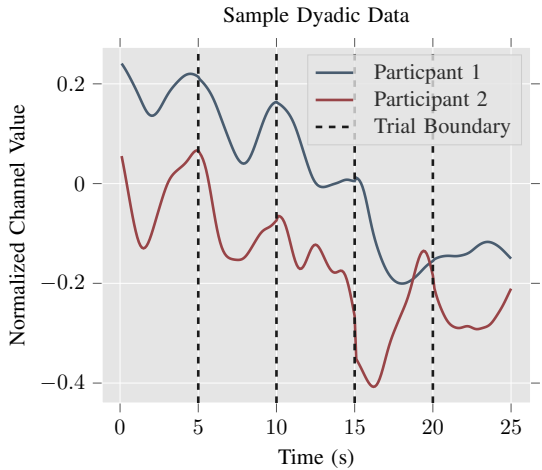}}
\caption{Example normalized channel data from two participants conducting five competitive trials.} \label{fig:dyadic_data}
\end{figure}

\Cref{fig:dyadic_data} shows sample data for one dyad over the course of five trials. Normalized single channel data from one participant is plotted against the matching channel data from their dyadic partner. The vertical red lines denote the end of a given trial and the start of a subsequent trial. As expected we see a delay in hemodynamic response of approximately 5--6 seconds \cite{cui2012nirs}.

\begin{table}\caption{\label{tab:dataset}Hyperscanning Dataset Summary}
\centering
\begin{tabular}{@{}lr@{}} \toprule
    \textbf{Quantity} & \textbf{Count}  \\
    \midrule
    \textbf{Total Trials} & $\mathbf{3,188}$ \\
    \midrule
    Cooperation Trials & $1,569$ \\
    Competition Trials & $1,619$ \\
    Total: & $3,188$ \\
    \midrule
    Male-Male Dyads & $1,675$ \\
    Female-Female Dyads & $1,513$ \\
    Total: & $3,188$ \\ \bottomrule
\end{tabular}
\end{table}

\begin{figure}[!htb]
\centering
\resizebox{0.4\textwidth}{!}{\input{figs/coop_reaction_times}}
\caption{Kernel density estimation of male-male and female-female cooperation trial reaction times.} \label{fig:coop_reaction_times}
\end{figure}  

\begin{figure}[!htb]
\centering
\resizebox{0.4\textwidth}{!}{\input{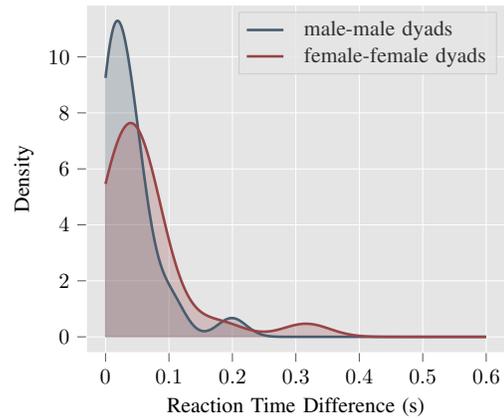}}
\caption{Kernel density estimation of male-male and female-female competition trial reaction times.} \label{fig:comp_reaction_times}
\end{figure} 

\subsection{Behavioral Measures}\label{behaviorial}
To provide further intuition about the behavioral component of the dyadic interactions, we briefly discuss differences in participant reaction time. Two participants that perfectly synchronize a button press event have a difference in reaction time of 0 seconds. \Cref{fig:coop_reaction_times} shows kernel density estimations (KDEs) of differences in reaction time for dyads completing cooperation tasks, while \cref{fig:comp_reaction_times} shows kernel density estimations of differences in reaction time for dyads completing competition tasks. Although overall trends remain strikingly similar across sex compositions, subtle differences emerge between male-male and female-female dyads. We ran the four non-parametric Mann-Whitney U tests described below to formally test for differences in underlying distributions:

\begin{enumerate}
    \item \textbf{Male-Male Test:} Given samples of reaction time differences for male-male dyads completing cooperation and competition tasks, test whether the underlying distributions are the same.
    \item \textbf{Female-Female Test:} Given samples of reaction time differences for female-female dyads completing cooperation and competition tasks, test whether the underlying distributions are the same.
    \item \textbf{Cooperation Test:} Given samples of reaction time differences for male-male and female-female dyads completing cooperation tasks, test whether the underlying distributions are the same. 
    \item \textbf{Competition Test:} Given samples of reaction time differences for male-male and female-female dyads completing competition tasks, test whether the underlying distributions are the same.
\end{enumerate}

\Cref{tab:mann_whit} summarizes the results. The cooperation test yields a p-value that is not statistically significant, while distinct trends emerge for both the competition and female-female tests. The only statistically significant result occurs during the male-male test. We seek to classify tasks based on patterns in cortical brain function rather than temporal differences in brain function and thus introduce CNN input data based on dynamic time warping. DTW identifies similar time series features that do not line up exactly in the time domain as discussed in \cref{cnn_input_data}.

\begin{table}\caption{\label{tab:mann_whit}Mann-Whitney U Test Results}
\centering
\begin{tabular}{@{}lcr@{}} \toprule
    \textbf{Test} & \textbf{T-statistic} & \textbf{P-value} \\
    \midrule
    Male-Male Test & $210.0$ & $0.0097$\\
    Female-Female Test & $203.0$ & $0.0900$\\
    \midrule
    Cooperation Test & $284.0$ & $0.3857$\\
    Competition Test & $225.0$ & $0.0696$\\ \bottomrule
\end{tabular}
\end{table}

\subsection{CNN Input Data} \label{cnn_input_data}
Convolutional neural networks are capable of extracting the salient features from input data with minimal \textit{a priori} intervention and feature engineering---a phenomenon that drives the ubiquitous use of CNNs in image recognition and computer vision tasks. We thus devote minimal attention to handcrafted feature engineering and instead test the proposed CNN-based classifier on channel-wise similarity scores computed using dynamic time warping.

\begin{figure}[!htb]
\centering
\resizebox{0.4\textwidth}{!}{\input{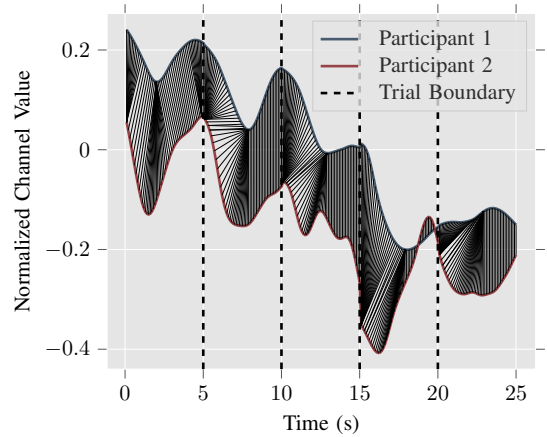}}
\caption{Optimal alignment of the time series data from the five competitive trials shown in Fig. \ref{fig:dyadic_data} as computed via dynamic time warping.} 
\label{fig:dtw_input_data}
\end{figure}

Dynamic time warping is a technique to find an optimal alignment between two sequences of time-series data. As the name suggests, the sequences are \textit{warped} nonlinearly to match each other \cite{muller2007dynamic}. Salient features in the data are then compared independent of nonlinearities in the time domain. The similarity score computed using DTW might thus more accurately describe how well two time series match each other when compared to a more conventional Euclidean distance measurement, as similar features will still be detected even if they do not line up exactly. \Citet{zhu2017dynamic} successfully employed DTW to average fNIRS signals and localize cortical brain activity. \cref{fig:dtw_input_data} shows how DTW can be used to align time series with a nonlinear time domain dependency. 


\cref{fig:dtw_input_data} presents the channel-wise similarity scores for a sample competition trial. Higher scores indicate a higher degree of similarity between the time series data.

\begin{figure}[!htb]
\centering
\resizebox{0.4\textwidth}{!}{\input{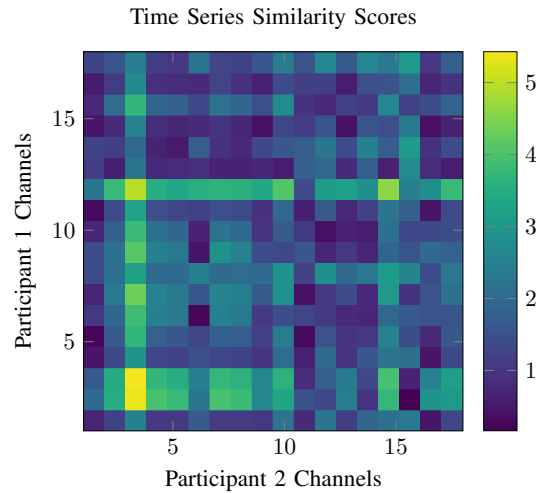}}
\caption{Channel-wise similarity score input data.} 
\label{fig:dtw_input_data_heatmap}
\end{figure} 

\section{Deep Learning Methods}

\subsection{LeNet-5 CNN}
Convolutional neural networks are a class of deep neural network that has achieved tremendous success in the object detection and image classification domains. Three defining characteristics of CNNs are \textit{local receptive fields}, \textit{weight replication}, and \textit{subsampling} \cite{lecun1998gradient}. 

Inputs are passed layer to layer in a CNN, and inputs at one layer are received from a small region of points---referred to as a \textit{local receptive field}---in the previous layer. Local connections allow a CNN to extract the most basic visual features such as edges and corners and then assemble the features in subsequent layers to extract higher order features. Furthermore, weight vectors are held identical for particular units with receptive fields scattered across the input image---a paradigm known as \textit{weight replication} (\cite{rumelhart1985learning, fukushima1982neocognitron, lecun1989generalization}, as cited in \cite{lecun1998gradient}). Units with identical weights are combined into a single plane and output a \textit{feature map}. This effectively shares feature detectors across the entire image. Finally, the concept of \textit{subsampling} is used to reduce the resolution of the feature maps and reduce the output sensitivity to distortions and feature shifts \cite{lecun1998gradient}. Subsampling can be implemented using an average pooling layer in TensorFlow.\footnote{https://www.tensorflow.org}

\begin{figure}[!htb]
\centering
\resizebox{0.48\textwidth}{!}{\input{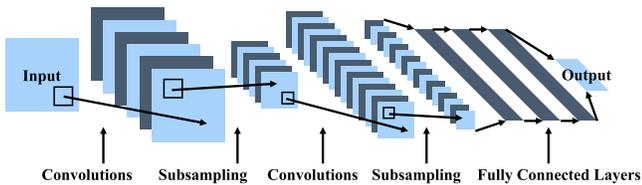}}
\caption{LeNet-5 CNN architecture.} \label{fig:lenet5}
\end{figure}

\Citet{lecun1998gradient} developed the popular CNN architecture referred to as LeNet-5 to incorporate the three previously discussed CNN design paradigms; a schematic is shown in \cref{fig:lenet5}. Two two-dimensional convolutions are performed on the input data, with intermediate subsampling layers reducing the resolution of the feature maps. The architecture concludes with fully connected layers leading to an output layer; the original LeNet-5 architecture proposed a convolutional layer fully connected to a single dense layer \cite{lecun1998gradient}. We employ a CNN based on the original LeNet-5 design to provide baseline results.

\subsection{Alternate CNN Architecture}

\begin{figure}[!htb]
\centering
\resizebox{0.48\textwidth}{!}{\input{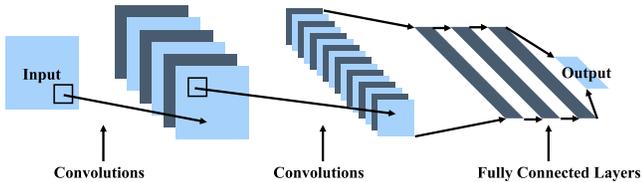}}
\caption{Proposed CNN architecture without average pooling layers.} \label{fig:proposed}
\end{figure}

The original LeNet-5 CNN uses average pooling layers to subsample the feature maps. We propose an alternate CNN architecture that closely mirrors the LeNet-5 scheme but does not contain the average pooling layers; an architectural diagram is presented in \cref{fig:proposed}. Removing the pooling layers theoretically heightens the output sensitivity to slight distortions in the feature map and maintains a high resolution throughout each layer. We suspect that the elevated resolution will enable the proposed CNN to extract nuanced features in the input fNIRS data and test this theory in the subsequent section.

\section{Results and Discussion}

We implemented a modified version of the LeNet-5 CNN in TensorFlow with two sets of alternating two-dimensional convolution and average pooling layers, followed by three dense layers with 128 units each. We also implemented our proposed CNN that does not include the average pooling layers. The architectures were tested on a series of four binary classification tasks:
\begin{enumerate}
    \item \textbf{MM Task Prediction:} Given male-male dyadic data, predict if task is cooperation or competition.
    \item \textbf{FF Task Prediction:} Given female-female dyadic data, predict if task is cooperation or competition.
    \item \textbf{Coop Sex Prediction:} Given cooperation task data, predict if dyad is male-male or female-female.
    \item \textbf{Comp Sex Prediction:} Given competition task data, predict if dyad is male-male or female-female.
\end{enumerate}

We conducted the four listed tests with the channel-wise similarity score input data on both CNN architectures. We used three-fold cross-validation to estimate the accuracy of each method. The CNNs were trained for 20 epochs with a batch size of 32 to balance the bias-error trade-off. \Cref{tab:dtw_results} show results for both CNN architectures. The code for the experiments, including the CNN implementations and the DTW similarity score dataset, is publicly available.\footnote{https://github.com/sisl/CNNHyperscanningClassification}

\begin{table}\caption{\label{tab:dtw_results}Results with DTW Input Data}
\centering
\begin{tabular}{@{}lrr@{}} \toprule
    \textbf{Classification Task} & \textbf{Pooling} & \textbf{Accuracy (\%)} \\ 
    \midrule
    MM Task Prediction & Yes & $55.86$ \\
    & No & $65.47$ \\    \midrule
    FF Task Prediction & Yes & $60.28$ \\
    & No & $66.08$ \\   \midrule
    Coop Sex Prediction & Yes & $75.80$ \\      
    & No & $80.61$ \\    \midrule
    Comp Sex Prediction & Yes & $72.84$ \\
    & No & $78.88$ \\ \bottomrule
\end{tabular}
\end{table}

The tabulated results indicate that dyadic sex composition can be classified more accurately than the task type. Furthermore, the results indicate that removing the pooling layers gives more predictive power to the CNN, as the heightened resolution at each convolutional layer likely improves feature extraction.
 
\begin{figure}[!h]
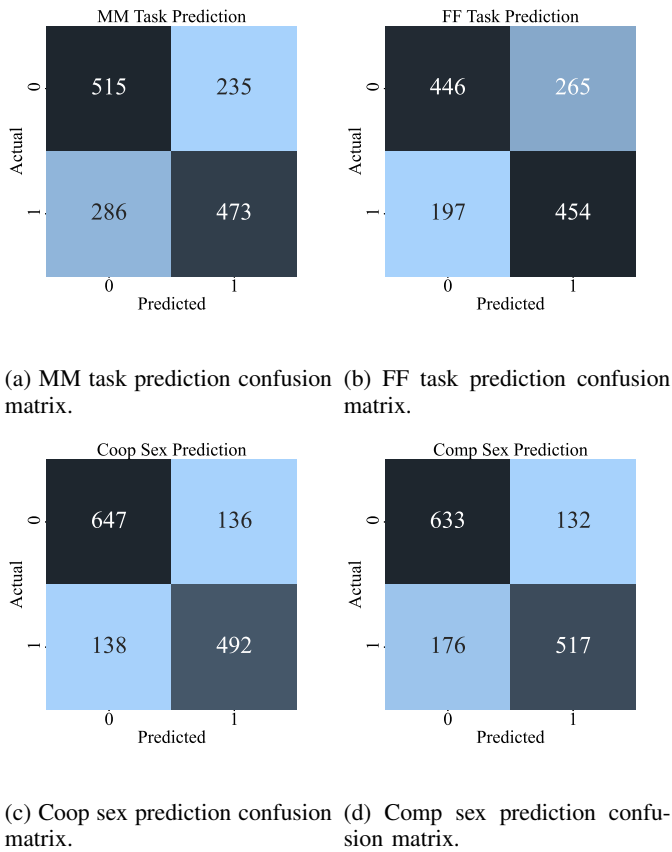

     \centering
     \begin{subfigure}[b]{0.24\textwidth}
\centering
\resizebox{1.0\textwidth}{!}{\input{figs/mm_task_pred_confusion}}
\smallskip
\caption{MM task prediction confusion matrix.}
\label{fig:mm_cm}
     \end{subfigure}
     \hfill
     \begin{subfigure}[b]{0.24\textwidth}
\centering
\resizebox{1.0\textwidth}{!}{\input{figs/ff_task_pred_confusion}}
\smallskip
\caption{FF task prediction confusion matrix.}
\label{fig:ff_cm}
     \end{subfigure}

     \begin{subfigure}[b]{0.24\textwidth}
\centering
\resizebox{1.0\textwidth}{!}{\input{figs/coop_gender_pred_confusion}}
\smallskip
\caption{Coop sex prediction confusion matrix.}
\label{fig:coop_cm}
     \end{subfigure}
     \hfill
     \begin{subfigure}[b]{0.24\textwidth}
\centering
\resizebox{1.0\textwidth}{!}{\input{figs/comp_gender_pred_confusion}}
\smallskip
\caption{Comp sex prediction confusion matrix.}
\label{fig:comp_cm}
     \end{subfigure}
        \caption{Confusion matrices for experiments with similarity score inputs and pooling layers removed.}
        \label{fig:genetic_result}
\end{figure}

 The classification accuracies obtained with similarity score inputs and pooling layers removed closely match the fNIRS binary classification state-of-the-art accuracies discussed in \cref{Related}. Confusion matrices displaying cumulative predictions across all cross-validation folds are shown in \cref{fig:mm_cm} to \cref{fig:comp_cm}. Only results for the experiments with pooling layers removed are displayed.

\section{Conclusion and Future Work}
In this work, we presented a CNN-based classifier for predicting the sex composition and task type of dyads performing cooperative and competitive tasks with fNIRS hyperscanning technology. We computed channel-wise similarity scores for each dyadic trial using dynamic time warping. We then used a convolutional neural network to classify fNIRS hyperscanning signals and obtained classification accuracies consistent with state-of-the-art accuracies on binary fNIRS classification tasks. This deep learning-based approach provides neuroimaging researchers with a novel strategy for analyzing dyadic inter-brain coherence.

The timely, accurate classification of fNIRS signals is critical in HHM scenarios wherein human actors are collaborating with robotic systems. Safety-critical HHM interactions occur across a wide range of high-stress environments including life-saving surgeries and high-risk aircraft and spacecraft landings such as pilot-induced oscillation events. Functional NIRS technology could be used to assess the social and mental workload of human participants and provide neurofeedback. Real-time BCI systems based on fNIRS have already been leveraged to conduct state estimation in flight simulators \cite{gateau2015real} and classify motor imagery \cite{hirsch2020online}.

In future work we will explore alternate CNN architectures and extend the binary classification tasks to multi-class classification tasks. Additional analyses on the channel-wise similarity scores could distill the most useful channels for classification tasks; such information could be incorporated in highly portable fNIRS low-channel solutions integrated into lightweight headsets such as the headband proposed in \citet{tsow2021low}. Furthermore, comprehensive deep learning model training on the neural signatures of a single human actor could learn to identify individual human behavior, promoting the accurate assessment of human identity and intentions in social interactions.

\section*{Acknowledgment}
This research was supported in part by the National Science Foundation Graduate Research Fellowship Program under Grant No. DGE-1656518 and the Kelvin Foundation.

\printbibliography

@article{lecun1998gradient,
  title={Gradient-based learning applied to document recognition},
  author={LeCun, Yann and Bottou, L{\'e}on and Bengio, Yoshua and Haffner, Patrick},
  journal={Proceedings of the IEEE},
  volume={86},
  number={11},
  pages={2278--2324},
  year={1998},
  publisher={Ieee}
}

@article{baker2016sex,
  title={Sex differences in neural and behavioral signatures of cooperation revealed by fNIRS hyperscanning},
  author={Baker, Joseph M and Liu, Ning and Cui, Xu and Vrticka, Pascal and Saggar, Manish and Hosseini, SM Hadi and Reiss, Allan L},
  journal={Scientific Reports},
  volume={6},
  number={1},
  pages={1--11},
  year={2016},
  publisher={Nature Publishing Group}
}

@article{muller2007dynamic,
  title={Dynamic time warping},
  author={M{\"u}ller, Meinard},
  journal={Information Retrieval for Music and Motion},
  pages={69--84},
  year={2007},
  publisher={Springer}
}

@article{peng2018single,
  title={Single-trial classification of f{NIRS} signals in four directions motor imagery tasks measured from prefrontal cortex},
  author={Peng, Hong and Chao, Jinlong and Wang, Sirui and Dang, Jie and Jiang, Fengqi and Hu, Bin and Majoe, Dennis},
  journal={IEEE Transactions on Nanobioscience},
  volume={17},
  number={3},
  pages={181--190},
  year={2018},
  publisher={IEEE}
}

@article{shin2014multiclass,
  title={Multiclass classification of hemodynamic responses for performance improvement of functional near-infrared spectroscopy-based brain--computer interface},
  author={Shin, Jaeyoung and Jeong, Jichai},
  journal={Journal of Biomedical Optics},
  volume={19},
  number={6},
  pages={$067009$},
  year={2014},
  publisher={International Society for Optics and Photonics}
}

@inproceedings{shamsi2019multi,
  title={Multi-class Classification of Motor Execution Tasks using f{NIRS}},
  author={Shamsi, F and Najafizadeh, L},
  booktitle={IEEE Signal Processing in Medicine and Biology Symposium (SPMB)},
  pages={1--5},
  year={2019}
  %organization={IEEE}
}

@article{sitaram2007temporal,
  title={Temporal classification of multichannel near-infrared spectroscopy signals of motor imagery for developing a brain--computer interface},
  author={Sitaram, Ranganatha and Zhang, Haihong and Guan, Cuntai and Thulasidas, Manoj and Hoshi, Yoko and Ishikawa, Akihiro and Shimizu, Koji and Birbaumer, Niels},
  journal={NeuroImage},
  volume={34},
  number={4},
  pages={1416--1427},
  year={2007},
  publisher={Elsevier}
}

@inproceedings{falk2010improving,
  title={Improving the performance of {NIRS}-based brain-computer interfaces in the presence of background auditory distractions},
  author={Falk, Tiago H and Paton, Kelly and Power, Sarah and Chau, Tom},
  booktitle={IEEE International Conference on Acoustics, Speech and Signal Processing},
  pages={517--520},
  year={2010}
  %organization={IEEE}
}

@article{power2010classification,
  title={Classification of prefrontal activity due to mental arithmetic and music imagery using hidden Markov models and frequency domain near-infrared spectroscopy},
  author={Power, Sarah D and Falk, Tiago H and Chau, Tom},
  journal={Journal of Neural Engineering},
  volume={7},
  number={2},
  pages={$026002$},
  year={2010},
  publisher={IOP Publishing}
}

@inproceedings{gottemukkula2011classification,
  title={Classification-guided feature selection for {NIRS}-based {BCI}},
  author={Gottemukkula, Vikas and Derakhshani, Reza},
  booktitle={IEEE/EMBS International Conference on Neural Engineering},
  pages={72--75},
  year={2011}
  %organization={IEEE}
}

@inproceedings{ayaz2009assessment,
  title={Assessment of cognitive neural correlates for a functional near infrared-based brain computer interface system},
  author={Ayaz, Hasan and Shewokis, Patricia A and Bunce, Scott and Schultheis, Maria and Onaral, Banu},
  booktitle={International Conference on Foundations of Augmented Cognition},
  pages={699--708},
  year={2009},
  organization={Springer}
}

@article{funane2011synchronous,
  title={Synchronous activity of two people's prefrontal cortices during a cooperative task measured by simultaneous near-infrared spectroscopy},
  author={Funane, Tsukasa and Kiguchi, Masashi and Atsumori, Hirokazu and Sato, Hiroki and Kubota, Kisou and Koizumi, Hideaki},
  journal={Journal of Biomedical Optics},
  volume={16},
  number={7},
  pages={077011},
  year={2011},
  publisher={International Society for Optics and Photonics}
}

@article{cui2012nirs,
  title={{NIRS}-based hyperscanning reveals increased interpersonal coherence in superior frontal cortex during cooperation},
  author={Cui, Xu and Bryant, Daniel M and Reiss, Allan L},
  journal={NeuroImage},
  volume={59},
  number={3},
  pages={2430--2437},
  year={2012},
  publisher={Elsevier}
}

@article{cheng2015synchronous,
  title={Synchronous brain activity during cooperative exchange depends on gender of partner: A fNIRS-based hyperscanning study},
  author={Cheng, Xiaojun and Li, Xianchun and Hu, Yi},
  journal={Human Brain Mapping},
  volume={36},
  number={6},
  pages={2039--2048},
  year={2015},
  publisher={Wiley Online Library}
}

@article{scholkmann2014review,
  title={A review on continuous wave functional near-infrared spectroscopy and imaging instrumentation and methodology},
  author={Scholkmann, Felix and Kleiser, Stefan and Metz, Andreas Jaakko and Zimmermann, Raphael and Pavia, Juan Mata and Wolf, Ursula and Wolf, Martin},
  journal={NeuroImage},
  volume={85},
  pages={6--27},
  year={2014},
  publisher={Elsevier}
}

@inproceedings{saadati2019convolutional,
  title={Convolutional neural network for hybrid f{NIRS}-{EEG} mental workload classification},
  author={Saadati, Marjan and Nelson, Jill and Ayaz, Hasan},
  booktitle={International Conference on Applied Human Factors and Ergonomics},
  pages={221--232},
  year={2019},
  organization={Springer}
}

@article{baker2017portable,
  title={Portable functional neuroimaging as an environmental epidemiology tool: a how-to guide for the use of f{NIRS} in field studies},
  author={Baker, Joseph M and Rojas-Valverde, Daniel and Guti{\'e}rrez, Randall and Winkler, Mirko and Fuhrimann, Samuel and Eskenazi, Brenda and Reiss, Allan L and Mora, Ana M},
  journal={Environmental Health Perspectives},
  volume={125},
  number={9},
  pages={$094502$},
  year={2017}
}

@inproceedings{asgher2020classification,
  title={Classification of mental workload ({MWL}) using support vector machines ({SVM}) and convolutional neural networks ({CNN})},
  author={Asgher, Umer and Khalil, Khurram and Ayaz, Yasar and Ahmad, Riaz and Khan, Muhammad Jawad},
  booktitle={International Conference on Computing, Mathematics and Engineering Technologies (iCoMET)},
  pages={1--6},
  year={2020},
  organization={IEEE}
}

@article{kocsis2006modified,
  title={The modified Beer--Lambert law revisited},
  author={Kocsis, Laszlo and Herman, Peter and Eke, Andras},
  journal={Physics in Medicine \& Biology},
  volume={51},
  number={5},
  pages={N91},
  year={2006},
  publisher={IOP Publishing}
}

@article{zhu2017dynamic,
  title={Dynamic time warping-based averaging framework for functional near-infrared spectroscopy brain imaging studies},
  author={Zhu, Li and Najafizadeh, Laleh},
  journal={Journal of Biomedical Optics},
  volume={22},
  number={6},
  pages={$066011$},
  year={2017},
  publisher={International Society for Optics and Photonics}
}

@techreport{rumelhart1985learning,
  title={Learning internal representations by error propagation},
  author={Rumelhart, David E and Hinton, Geoffrey E and Williams, Ronald J},
  year={1985},
  institution={California Univ San Diego La Jolla Inst for Cognitive Science}
}

@article{fukushima1982neocognitron,
  title={Neocognitron: A new algorithm for pattern recognition tolerant of deformations and shifts in position},
  author={Fukushima, Kunihiko and Miyake, Sei},
  journal={Pattern Recognition},
  volume={15},
  number={6},
  pages={455--469},
  year={1982},
  publisher={Elsevier}
}

@article{lecun1989generalization,
  title={Generalization and network design strategies},
  author={LeCun, Yann and others},
  journal={Connectionism in Perspective},
  volume={19},
  pages={143--155},
  year={1989},
  publisher={Elsevier Zurich, Switzerland}
}

@article{di2019brain,
  title={Brain--computer interface-based adaptive automation to prevent out-of-the-loop phenomenon in air traffic controllers dealing with highly automated systems},
  author={Di Flumeri, Gianluca and De Crescenzio, Francesca and Berberian, Bruno and Ohneiser, Oliver and Kramer, Jan and Aric{\`o}, Pietro and Borghini, Gianluca and Babiloni, Fabio and Bagassi, Sara and Piastra, Sergio},
  journal={Frontiers in Human Neuroscience},
  volume={13},
  pages={296},
  year={2019},
  publisher={Frontiers}
}

@article{balters2020capturing,
  title={Capturing human interaction in the virtual age: a perspective on the future of f{NIRS} hyperscanning},
  author={Balters, Stephanie and Baker, Joseph M and Hawthorne, Grace and Reiss, Allan L},
  journal={Frontiers in Human Neuroscience},
  volume={14},
  pages={458},
  year={2020},
  publisher={Frontiers}
}

@inproceedings{mehta2019real,
  title={Real-time driver drowsiness detection system using eye aspect ratio and eye closure ratio},
  author={Mehta, Sukrit and Dadhich, Sharad and Gumber, Sahil and Jadhav Bhatt, Arpita},
  booktitle={International Conference on Sustainable Computing in Science, Technology and Management (SUSCOM)},
  year={2019}
}

@inproceedings{jung2015using,
  title={Using robots to moderate team conflict: the case of repairing violations},
  author={Jung, Malte F and Martelaro, Nikolas and Hinds, Pamela J},
  booktitle={ACM/IEEE International Conference on Human-Robot Interaction},
  pages={229--236},
  year={2015}
}

@article{de2020towards,
  title={Towards a theory of longitudinal trust calibration in human--robot teams},
  author={De Visser, Ewart J and Peeters, Marieke MM and Jung, Malte F and Kohn, Spencer and Shaw, Tyler H and Pak, Richard and Neerincx, Mark A},
  journal={International Journal of Social Robotics},
  volume={12},
  number={2},
  pages={459--478},
  year={2020},
  publisher={Springer}
}

@article{ferrari2012brief,
  title={A brief review on the history of human functional near-infrared spectroscopy (f{NIRS}) development and fields of application},
  author={Ferrari, Marco and Quaresima, Valentina},
  journal={NeuroImage},
  volume={63},
  number={2},
  pages={921--935},
  year={2012},
  publisher={Elsevier}
}

@article{plichta2006event,
  title={Event-related functional near-infrared spectroscopy (f{NIRS}): are the measurements reliable?},
  author={Plichta, Michael M and Herrmann, Martin J and Baehne, CG and Ehlis, A-C and Richter, MM and Pauli, Paul and Fallgatter, Andreas J},
  journal={NeuroImage},
  volume={31},
  number={1},
  pages={116--124},
  year={2006},
  publisher={Elsevier}
}

@article{molavi2012wavelet,
  title={Wavelet-based motion artifact removal for functional near-infrared spectroscopy},
  author={Molavi, Behnam and Dumont, Guy A},
  journal={Physiological Measurement},
  volume={33},
  number={2},
  pages={259},
  year={2012},
  publisher={IOP Publishing}
}

@article{liu2016nirs,
  title={{NIRS}-based hyperscanning reveals inter-brain neural synchronization during cooperative Jenga game with face-to-face communication},
  author={Liu, Ning and Mok, Charis and Witt, Emily E and Pradhan, Anjali H and Chen, Jingyuan E and Reiss, Allan L},
  journal={Frontiers in Human Neuroscience},
  volume={10},
  pages={82},
  year={2016},
  publisher={Frontiers}
}

@article{huppert2009homer,
  title={HomER: a review of time-series analysis methods for near-infrared spectroscopy of the brain},
  author={Huppert, Theodore J and Diamond, Solomon G and Franceschini, Maria A and Boas, David A},
  journal={Applied Optics},
  volume={48},
  number={10},
  pages={D280--D298},
  year={2009},
  publisher={Optical Society of America}
}

@article{cui2011quantitative,
  title={A quantitative comparison of {NIRS} and f{MRI} across multiple cognitive tasks},
  author={Cui, Xu and Bray, Signe and Bryant, Daniel M and Glover, Gary H and Reiss, Allan L},
  journal={NeuroImage},
  volume={54},
  number={4},
  pages={2808--2821},
  year={2011},
  publisher={Elsevier}
}

@article{miller2019inter,
  title={Inter-brain synchrony in mother-child dyads during cooperation: an fNIRS hyperscanning study},
  author={Miller, Jonas G and Vrti{\v{c}}ka, Pascal and Cui, Xu and Shrestha, Sharon and Hosseini, SM Hadi and Baker, Joseph M and Reiss, Allan L},
  journal={Neuropsychologia},
  volume={124},
  pages={117--124},
  year={2019},
  publisher={Elsevier}
}

@article{mayseless2019real,
  title={Real-life creative problem solving in teams: f{NIRS} based hyperscanning study},
  author={Mayseless, Naama and Hawthorne, Grace and Reiss, Allan L},
  journal={NeuroImage},
  volume={203},
  pages={116161},
  year={2019},
  publisher={Elsevier}
}

@article{liu2017inter,
  title={Inter-brain network underlying turn-based cooperation and competition: A hyperscanning study using near-infrared spectroscopy},
  author={Liu, Tao and Saito, Godai and Lin, Chenhui and Saito, Hirofumi},
  journal={Scientific Reports},
  volume={7},
  number={1},
  pages={1--12},
  year={2017},
  publisher={Nature Publishing Group}
}

@article{zhang2017gender,
  title={Gender difference in spontaneous deception: a hyperscanning study using functional near-infrared spectroscopy},
  author={Zhang, Mingming and Liu, Tao and Pelowski, Matthew and Yu, Dongchuan},
  journal={Scientific Reports},
  volume={7},
  number={1},
  pages={1--13},
  year={2017},
  publisher={Nature Publishing Group}
}

@article{zhang2017social,
  title={Social risky decision-making reveals gender differences in the {TPJ}: A hyperscanning study using functional near-infrared spectroscopy},
  author={Zhang, Mingming and Liu, Tao and Pelowski, Matthew and Jia, Huibin and Yu, Dongchuan},
  journal={Brain and Cognition},
  volume={119},
  pages={54--63},
  year={2017},
  publisher={Elsevier}
}

@article{lu2020gender,
  title={Gender of partner affects the interaction pattern during group creative idea generation},
  author={Lu, Kelong and Teng, Jing and Hao, Ning},
  journal={Experimental Brain Research},
  volume={238},
  number={5},
  pages={1157--1168},
  year={2020},
  publisher={Springer}
}

@article{gateau2015real,
  title={Real-time state estimation in a flight simulator using fNIRS},
  author={Gateau, Thibault and Durantin, Gautier and Lancelot, Francois and Scannella, Sebastien and Dehais, Frederic},
  journal={PLOS One},
  volume={10},
  number={3},
  pages={e0121279},
  year={2015},
  publisher={Public Library of Science San Francisco, CA USA}
}

@inproceedings{hirsch2020online,
  title={Online Classification of Motor Imagery Using EEG and fNIRS: A Hybrid Approach with Real Time Human-Computer Interaction},
  author={Hirsch, Gerald and Dirodi, Matilde and Xu, Ren and Reitner, Patrick and Guger, Christoph},
  booktitle={International Conference on Human-Computer Interaction},
  pages={231--238},
  year={2020},
  organization={Springer}
}

@article{tsow2021low,
  title={A low-cost, wearable, do-it-yourself functional near-infrared spectroscopy ({DIY}-f{NIRS}) headband},
  author={Tsow, Francis and Kumar, Anupam and Hosseini, SM Hadi and Bowden, Audrey},
  journal={HardwareX},
  volume={10},
  pages={e00204},
  year={2021},
  publisher={Elsevier}
}

\end{document}